\documentclass[final,twocolumn]{svjour3}
\usepackage{graphicx}
\usepackage{balance}
\usepackage{multicol}
\journalname{Astrophysics and Space Science}

\begin{document}

\title{IC 4406: a radio-infrared comparison}
\author{Luciano Cerrigone         \and
        Joseph L. Hora \and
        Grazia Umana   \and
        Corrado Trigilio
}
\institute{L. Cerrigone and J. Hora \at
              Harvard-Smithsonian Center for Astrophysics \\
              60 Garden St \\
		Cambridge, MA, USA \\
           \and
           L. Cerrigone \at
              Tel.: +1-617-495-7182\\
              Fax: +1-617-496-7490\\
\email{lcerrigone@cfa.harvard.edu} \\
		\and
		G. Umana and C. Trigilio \at
		Osservatorio Astrofico Catania \\
		via S. Sofia 78 \\
		Catania, Italy \\
}
\date{November, 13-17 2006}
\maketitle
\begin{abstract}
IC 4406 is a large (about 100$'' \times 30''$) southern bipolar planetary nebula, composed by two elongated lobes, extending from a
bright central region, where there is evidence for the presence of a large torus of gas and dust.
In this poster we show new observations of this source performed with IRAC (Spitzer Space Telescope) and the Australia Telescope
Compact Array. Although the possibility for faint extended emission to be missing in the radio maps cannot be ruled out, flux from the
ionized gas appears to be concentrated in the bright central region. Comparing ATCA to IRAC images, it seems that, like in other planetary
nebulae, ionized and neutral components spatially co-exist in IC 4406.
\end{abstract}
\section{Introduction}
\label{intro}
IC4406 is a well studied southern planetary nebula. It has been imaged with several telescopes and at different wavelength ranges. It shows two H$_2$ lobes \cite{storey},  orthogonal to the nebula's major axis and each 15$''$ away from the center. These peaks are approximately coincident with the two blobs observed in H$\alpha$+NII and OIII \cite{sahai}, interpreted as indicative of the presence of a dense equatorial torus of dust. CO maps \cite{sahai} show the presence of a collimated high velocity outflow in the polar direction. Hubble images in NII, H$\alpha$ and OIII have revealed the existence of an intricate system of dark lane features \cite{odell}, which led to the name of \lq\lq Retina Nebula\rq\rq~for this object \cite{odell}.\\
We have observed this source in the radio range to inspect the distribution of the ionized gas in its envelope and in the infrared to check for emission from the equatorial dust and molecular gas.

\section{Observations and Data reduction}
\label{sec1}
Radio observations were performed at the {\bf A}ustralia {\bf T}e\-le\-scope {\bf C}ompact {\bf A}rray on November 24, 2005 (UT: from 17:00:00 to 08:00:00) and December 11, 2005 (UT: from 15:30:00 to 02:00:00), simultaneously at 4.8 and 8.6 GHz.
The November run was performed with the array in 1.5C configuration, while for the December one the configuration
was 6.0A. The adopted configurations are both linear but with different antenna positions, giving maximum baselines of
4500 m (1.5C) and 5939 m (6.0A), minimum baselines of 77 m (1.5C) and 337 m (6.0A). The pre-calibration of the
array was performed observing 0823-500, while the absolute flux calibrator was 1934-638. Another target was also
observed during our two runs and the total on-target time was about 7 hours for each of the two. The phase calibrator
chosen for IC 4406 was 1431-48, which is 4.76$^{\circ}$ away from the target. The data were reduced with {\bf MIRIAD}, following a
standard reduction procedure as recommended in the MIRIAD User's Guide. The data from the two runs were
combined into one dataset, obtaining a {\it uv} coverage from 0.9 to 96 k$\lambda$ at 4.8 GHz and from 1.5 to 172 k$\lambda$ at 8.6 GHz.
These correspond to an angular resolution of 2.2$''$ at 4.8 GHz and 1.2$''$ at 8.6 GHz, while the largest observable
structures (Largest Angular Scale) are 230$''$ and 140$''$ respectively. 

Infrared observations were performed with the {\bf I}nfra\-{\bf R}ed {\bf A}rray {\bf C}amera onboard Spitzer Space Te\-le\-scope at 3.6, 4.5,
5.8 and 8.0~$\mu m$ on March 06, 2004 (UT 09:54:16.311). Six High Dynamic Range 30 sec dithered frames were obtained
at each wavelength, for a total exposure time of 180 sec per channel. In Tab.\ref{tab:pointing} the pointing coordinates for the two datasets are listed.

\begin{table}
\centering
{\footnotesize
\caption{J2000 pointing coordinates for our target and its phase calibrator, used for our ATCA run, are shown.}
\label{tab:pointing}
\begin{tabular}{lcc}
\hline
 \noalign{\smallskip}
  & {\it RA} & {\it DEC}  \\
{\bf SST} & &  \\
IC 4406 & 14:22:27.66 & -44:09:00.0 \\
\noalign{\smallskip}
\hline
\noalign{\smallskip}
{\bf ATCA} & & \\
IC 4406 & 14:22:26.28 & -44:09:00.0  \\
1431-48 & 14:35:16.80 & -48:21:47.76  \\
\hline
\end{tabular}}
\end{table}
Ba\-sic Ca\-libration Data were retrieved from Spitzer ar\-chive, cleaned to correct such artifacts as mux-bleeding and
banding and then coadded using {\bf IRACproc} \cite{schuster}.
The whole nebula in each image was boxed with a polygon and the flux within such region summed up. The same
procedure was adopted for the field star observed West of the central core, so that its flux was subtracted to the overall
nebula's one. The result was corrected for an infinite aperture, according to the IRAC Data Handbook and then
converted into mJy (SST BCD files are in units of MJy/sr). \\
Tab.\ref{tab:fluxes} summarizes the results of our radio and infrared observations.
\begin{table}
\centering
{\footnotesize
\caption{In lines one and two the wavelengths and relative flux measurements from Spitzer Space Telescope are listed. Lines 3--5 summarize the radio observations: frequencies and relative flux values in columns 3 and 4, the rms in column 5 (the latter is the same within errors for both frequencies). Flux unit is mJy.}
\begin{tabular}{lcccc}
\noalign{\smallskip}\hline
\noalign{\smallskip}
{\bf SST} & {\it 3.6} $\mu m$ & {\it 4.5} $\mu m$ & {\it 5.8} $\mu m$ & {\it 8.0} $\mu m$\\
 \noalign{\smallskip}
IC 4406 &   62.7 & 110.0 & 132.4 & 307.2 \\
\noalign{\smallskip}
\hline
 \noalign{\smallskip}
 {\bf ATCA} & & {\it 4.8 GHz} & {\it 8.6 GHz} & $\sigma$ \\
 \noalign{\smallskip}
IC 4406 & & 96.47 & 91.15 & 0.04 \\
1431-48 & &  1020 & 650 & 1 \\
\noalign{\smallskip}\hline
\end{tabular}}
\label{tab:fluxes}
\end{table}

\section{Results and Discussion}
\label{sec2}
\begin{figure}
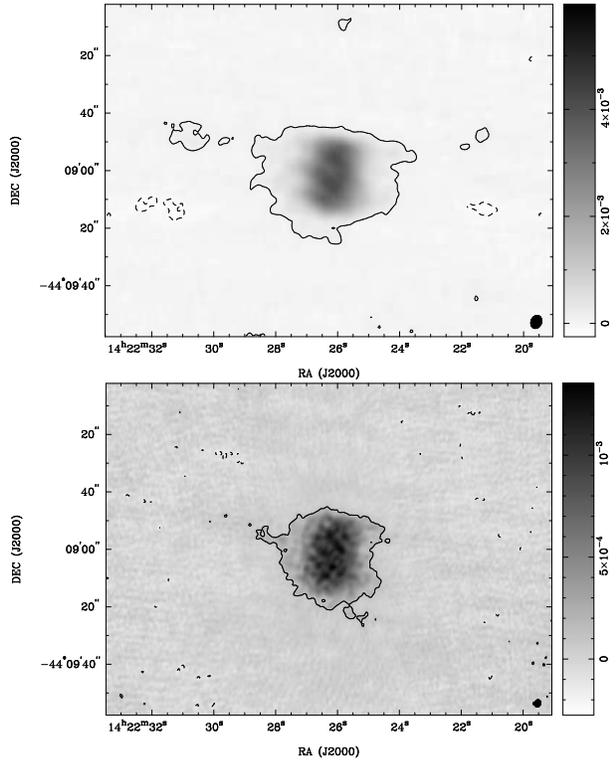

\centering
  \includegraphics[width=5cm,angle=270]{ic4406.Cnat.ps}
 \includegraphics[width=5cm,angle=270]{ic4406.Xnat.ps}
\caption{ Radio maps of IC 4406 ({\it top}: 4.8, {\it bottom}: 8.6 GHz). The shown contours indicate the $\pm3\sigma$ levels. Both maps were produced with NATURAL weights. The synthetic beam is shoiwn in the bottom right of each map and the flux density unit is Jy/beam.}
\label{fig:radiomaps}       
\end{figure}

\begin{figure}
\centering
  \includegraphics[width=8cm]{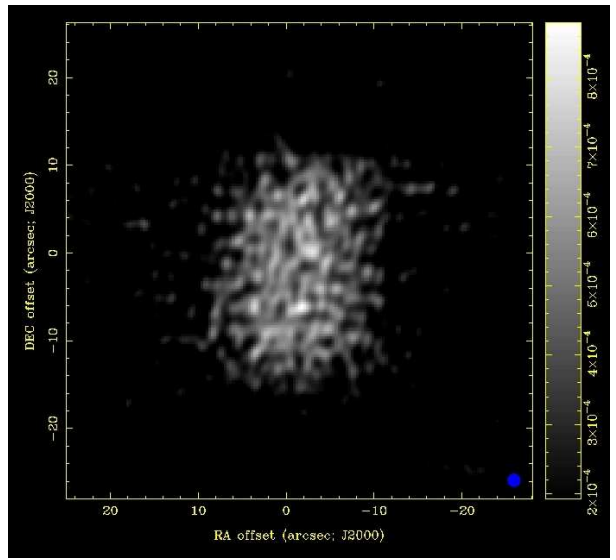}
\caption{Zoom of the 8.6 GHz map. This map was obtained with ROBUST weighting, to balance better resolution and sensitivity to faint structures. The RESTORed map was scaled to 1.5$''$ beam, shown in the bottom right. The flux density is in Jy/beam and the plot ranges from the 3$\sigma$ level to the peak value.}
\label{fig:zoom}       
\end{figure}

\begin{figure}
\centering
\includegraphics[width=8cm]{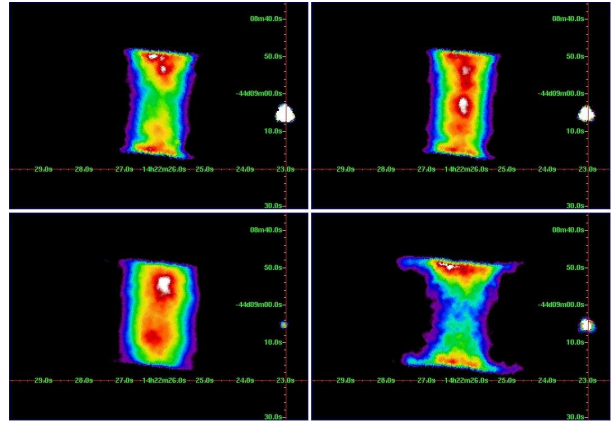}
\caption{Zooms of the central core emission observed with IRAC.
The scale is linear: maximum is the emission peak in the area,
minimum is 50\% of the maximum. Clockwise from top left: 3.6,
4.5, 5.8, 8.0 $\mu m$.}
\label{fig:composite}       
\end{figure}

\begin{figure}
\centering
  \includegraphics[width=8cm]{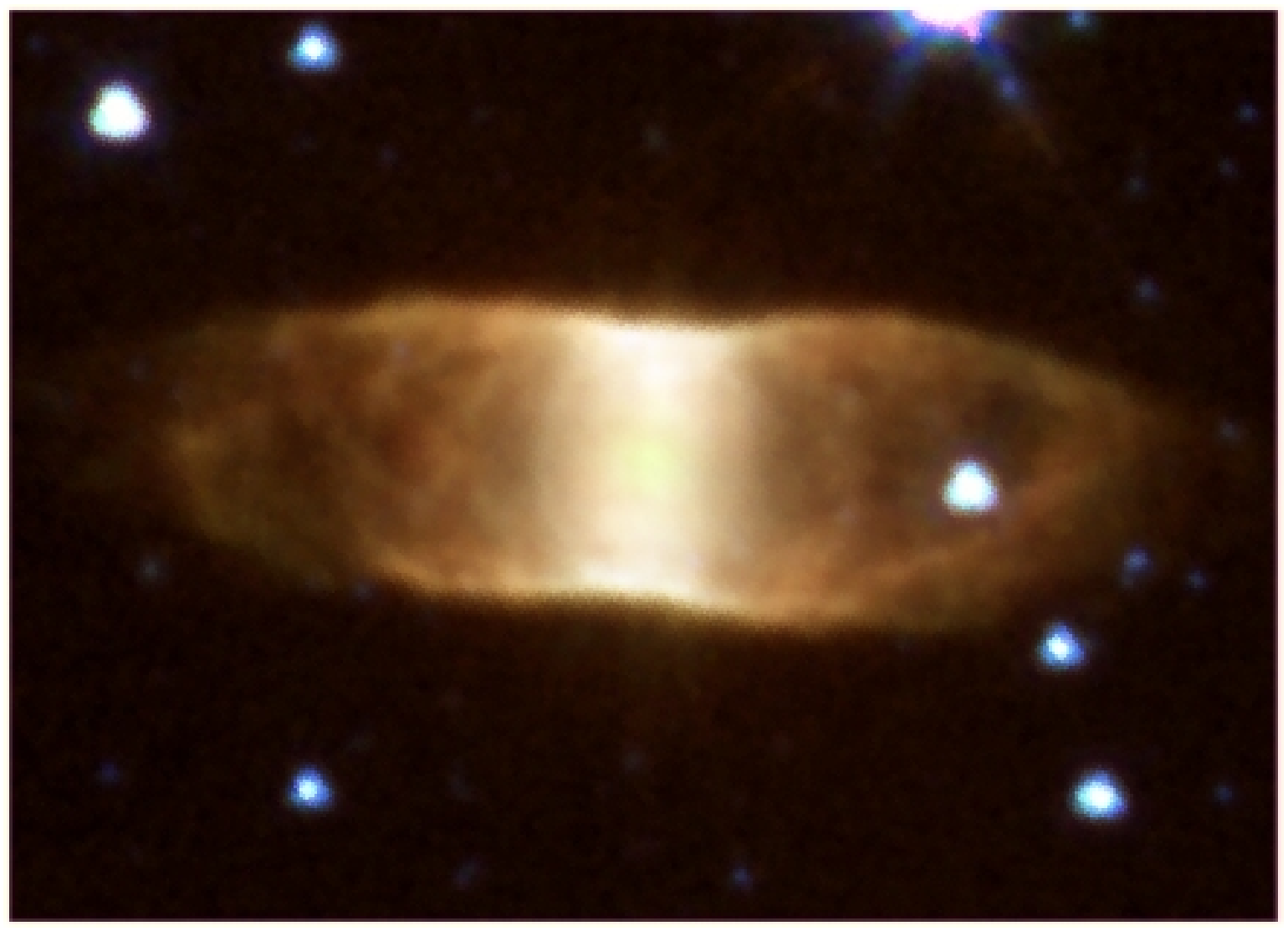}
\includegraphics[width=8cm]{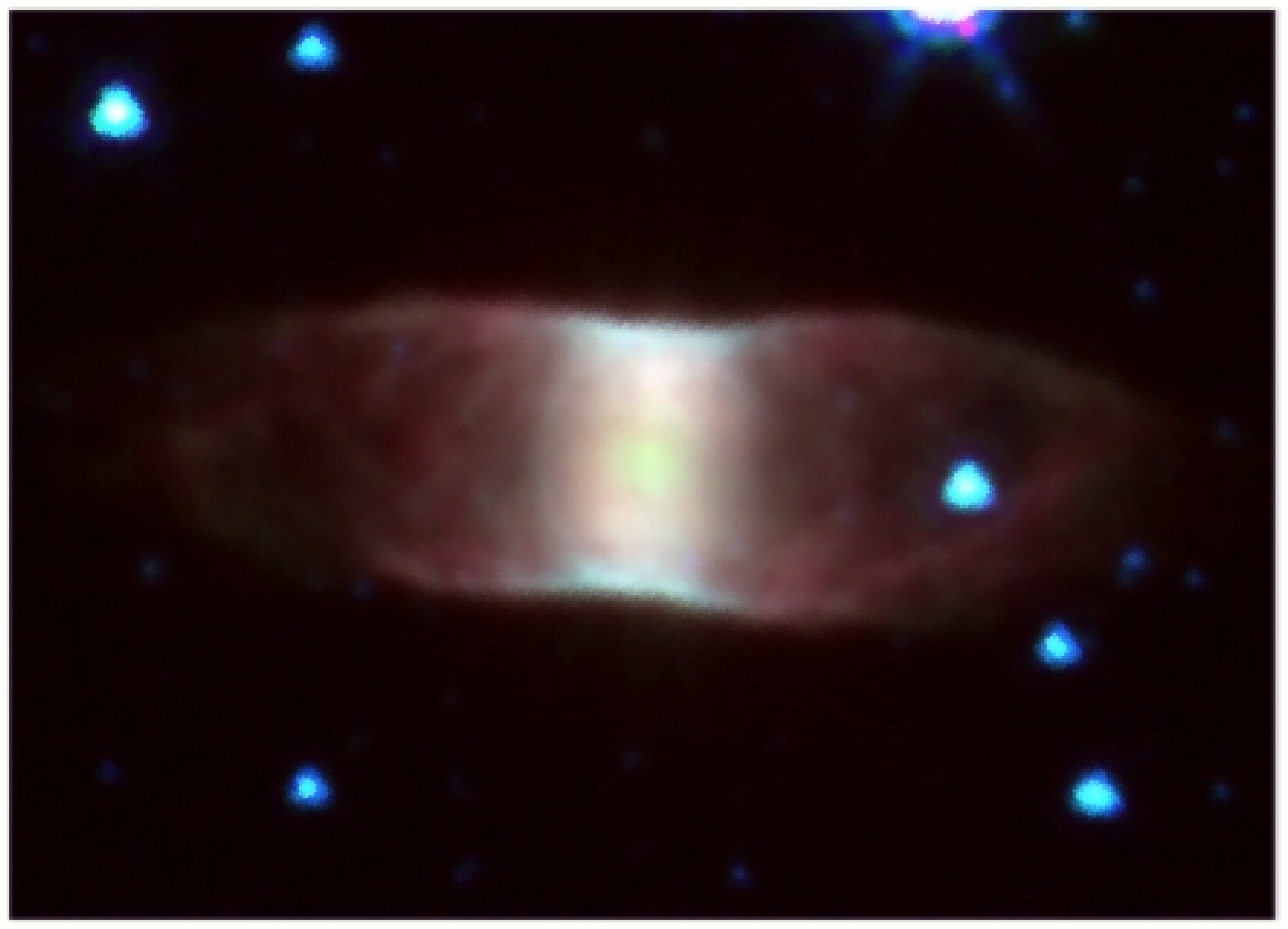}
 \includegraphics[width=8cm]{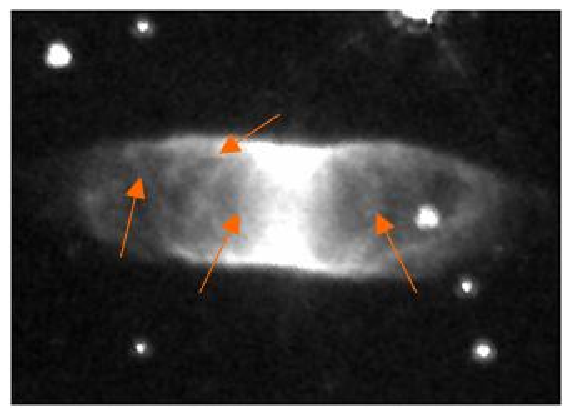}
\caption{Images obtained with IRAC. From top to bottom: overlay of IRAC channels (blue is 3.6, green 4.5, orange 5.8 and red is 8.0 $\mu$m); as in the previous image but not including 5.8 $\mu$m channel, for better viewing the central core (torus) emission; 5.8 $\mu$m image scaled for a better view of the arch structures mentioned in the text, some of which are pointed by the arrows.}
\label{fig:irac}       
\end{figure}

Our radio maps (Fig.\ref{fig:radiomaps}) show the presence at 4.8 GHz of a 42$''\times 56''$ (3$\sigma$ level) emitting region, elongated in E-W direction; at
a 10\% of the peak level the size of the emitting region is restricted to 32$'' \times 32''$. At 8.6 GHz the 10\% of the peak level
gives a size of 36$'' \times 40''$, in agreement with its 3$\sigma$ level size.
The maps don't show any N-S blobs of emission. What is shown is a very intricate system of emitting lanes (Fig.\ref{fig:zoom}),
which resembles what is observed by Hubble Space Telescope \cite{odell}.
An inspection of the amplitude vs. {\it uv}-distance plots leads to the conclusion that faint extended emission may be
missing in our radio maps despite the combination of an extended and a compact array configuration.
Fig.\ref{fig:composite} shows IRAC images of the central equatorial area of the nebula. They were plotted with a linear scale having the
peak flux and 50\% of it as thresholds. Channel 4 is clearly showing the emission from the torus of dust surrounding
the central star. Its size is about 28$'' \times 20''$, elongated in N-S direction and the angular distance between its peaks is
about 14$''$. The overall size of the torus matches with the approximate size of the nebula in N-S (30$''$), although its
peaks are much closer to the center than the H$_2$ blobs repoted in \cite{storey}, which can indicate the torus is partly shielding the molecular
gas from UV radiation.
Fig.\ref{fig:irac} is a combination of IRAC channels in linear scale. Despite the lower
resolution compared to Hubble images, IRAC is able to detect the faint emission from the neutral components in the
envelope and reveals the structure of the elongated lobes. They show arches at different distances and inclinations
from the central star, connected to the mass loss history. The arches which are angularly closer to the central star
show larger blue emission. This implies they have higher temperature, which can be explained if they are intrinsically
closer to the star. The overall structure observed in the envelope corresponds to the idea of the central torus as a main collimating agent.

\begin{figure}
\hspace{-1cm}
\includegraphics[width=9cm]{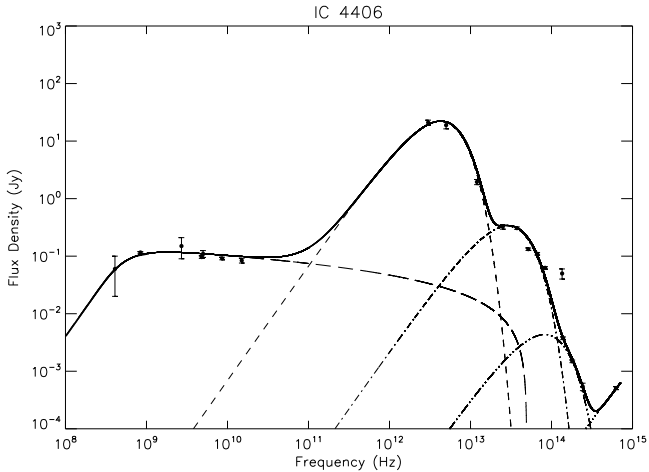}
\caption{Spectral energy distribution with literature (solid circles) and new (open squares, with errors within each square) data. The single contributions are also plotted. Long dashed curve: free-free; dashed: blackbody at 72 K; dash-dot: blackbody at 490 K; dash-dot-dot: blackbody at 1400 K; dots: blackbody at 10$^5$ K. }
\label{fig:sed}
\end{figure}

Collecting literature data and adding our new measurements, we have built up the SED for our target from the radio to
the near-IR. We have overplotted to the data the sum of the theoretical curves for free-free emission and blackbody
radiation. Three blackbodies with different temperatures were used to match the infrared data. The radio free-free emission was reproduced according to the equations in \cite{pott1}. One more blackbody was plotted to match one optical point, to correct near-IR emission at shorter wavelengths. Tab.\ref{tab:match} summarizes the parameters used for the plot. 
\begin{table}
\centering
{\footnotesize
\caption{Parameters adopted to match observational points in Fig.\ref{fig:sed}. Emission measure is in cm$^{-6} \cdot$ pc, temperatures in Kelvin.}
\label{tab:match}
\begin{tabular}{cccc}
\noalign{\smallskip}\hline
 \noalign{\smallskip}
{\it IRAS} & {\it IRAC} & {\it 2MASS} & {\it Central star} \\
T=72 & T=490 & T=1400 & T=10$^5$ \\
\noalign{\smallskip}\hline
 & & & \\
\multicolumn{4}{c}{{\it Radio}}\\
\multicolumn{2}{c}{EM=9.696 $\cdot$ 10$^5$} &  \multicolumn{2}{c}{T=10000} \\
\noalign{\smallskip}\hline
\end{tabular}}
\newpage
\end{table}
\balance
The curves were separately normalized to one point before being summed up to match the data points (Fig.\ref{fig:sed}). As reported in \cite{sahai}, more than one dust component is necessary to match the observations. Since fewer points were used in that study, a different number of components was found. Although the possibility of $H_2$ line emission in IRAC bands cannot be ruled out, thus changing the continuum shape, according to our calculations, cold,
warm and hot dust need to exist in the circumstellar environment to explain current observations. Tab.\ref{tab:results} reports some parameters calculated
assuming two different distances as reported in the literature (\cite{sahai}, \cite{odell}).
\begin{table}
\centering
{\footnotesize
\caption{Parameters calculated for two reported distances to our target. Density and ionized mass are calculated from 5 GHz data, according to \cite{pott1}, dust mass from 25 $\mu m$ data, according to \cite{pott2}.}
\label{tab:results}
\begin{tabular}{lcc}
\noalign{\smallskip}\hline
\noalign{\smallskip}
Distance (kpc) & 0.7 & 1.6 \\
M$_{ionized}$ ($M_\odot$) & 0.11 & 0.88 \\
M$_{dust}$  (10$^{-5}$ $M_\odot$) & 4.54 & 23.7 \\
M$_D$/M$_{ion}$ & 4.06 $\cdot$ 10$^{-4}$ & 2.68 $\cdot$ 10$^{-4}$ \\
Density ($cm^{-3}$) & 1.37 $\cdot$ 10$^3$ & 0.90 $\cdot$ 10$^3$ \\
\noalign{\smallskip}\hline
\end{tabular}}
\end{table}

We can speculate that in such a diversified dust environment further lower temperature components may exist and
high sensitivity, high angular resolution observations with ALMA will sure give a fundamental contribution to
understand the physics of circumstellar envelopes in planetary nebulae.

\begin{acknowledgements}
L. Cerrigone acknowledges the Smithsonian Astrophysical Observatory for funding through the SAO Predoctoral Fellowship Program.\\
This work is based in part on observations made with the Spitzer Space Telescope, operated by Jet Propulsion Laboratory under NASA contract 1407. \\
The Australia Telescope Compact Array is part of the Australia Telescope which is funded by the Commonwealth of Australia for operation as a National Facility managed by CSIRO. \\
\end{acknowledgements}

%
%

\end{document}